\documentclass[12pt]{article}
\usepackage{fullpage}
\usepackage{color,graphicx}

\usepackage{amsmath,verbatim,amssymb,epsfig}
\usepackage{smartref,color}
\usepackage{longtable}
\usepackage{rotating}
\usepackage{graphicx}
\usepackage{subfig}
\usepackage{setspace}
\usepackage{booktabs}
\usepackage{xfrac}
\usepackage{multirow}
\usepackage{cite}
\usepackage{color}

\begin{document}
	\baselineskip=24pt
\begin{center}
	{\Large \bf Elastic Priors to Dynamically Borrow Information from Historical Data in Clinical Trials}
\end{center}

\vspace{2mm}
\begin{center}
	Liyun Jiang$^{1, 3}$, Lei Nie$^2$, and Ying Yuan$^1$\\
\end{center}
\vskip 0.1in\noindent

\noindent{$^1$Department of Biostatistics,
The University of Texas MD Anderson Cancer Center, Houston, TX\\
$^2$ Center for Drug Evaluation and Research, Food and Drug Administration (FDA), Silver Spring, MD\\
$^3$ Research Center of Biostatistics and Computational Pharmacy, China Pharmaceutical University, Nanjing, China.

\vspace{6mm}

\noindent \emph{\textbf{Abstract}}:
Use of historical data and real-world evidence holds great potential to improve the efficiency of clinical trials. One major challenge is how to effectively borrow information from historical data while maintaining a reasonable type I error.  We propose the elastic prior approach to address this challenge and achieve dynamic information borrowing.  Unlike existing approaches, this method proactively controls the behavior of dynamic information borrowing and type I errors  by incorporating a well-known concept of clinically significant difference through an elastic function, defined as a monotonic function of a congruence measure between historical data and trial data. The elastic function is constructed to satisfy a set of information-borrowing constraints prespecified by researchers or regulatory agencies, such that the prior will borrow information when historical and trial data are congruent, but refrain from information borrowing when historical and trial data are incongruent.  In doing so, the elastic prior improves power and reduces the risk of data dredging and bias. The elastic prior is information borrowing consistent, i.e. asymptotically controls type I and II errors at the nominal values when historical data and trial data are not congruent, a unique characteristics of the elastic prior approach.  Our simulation study that evaluates the finite sample characteristic confirms that, compared to existing methods, the elastic prior has better type I error control and yields competitive or higher power. 

\par\vspace{3mm}

\section{Introduction}
\label{s:intro}

Real-world data (RWD) or evidence plays an increasingly important role in health care decisions. The 21st Century Cures Act, signed into law in 2016, emphasizes modernization of clinical trial designs, including the use of real-world evidence to support approval of new indications for approved drugs or to satisfy post-approval study requirements. The FDA released related guidance in the ``Use of Real-World Evidence to Support Regulatory Decision-Making for Medical Devices" in 2017 (FDA 2017), and a draft guidance on ``Submitting Documents Using Real-World Data and Real-World Evidence to FDA for Drugs and Biologics Guidance for Industry" in 2019 (FDA 2019).

Use of RWD to facilitate medical decisions is an extremely broad topic. We here focus on the use of historical data $D_h$ to improve the efficiency and guide decision making of randomized controlled trials (RCTs). For ease of exposition, we assume two-arm RCTs and historical data are only available on the control. It is straightforward to extend the proposed methodology to multiple-arm RCTs and to cases where historical data are also available for the treatment arm.  The question of interest is how to leverage information from $D_h$  to increase the power of comparing the treatment efficacy between the control and treatment arms. This problem is also known as {augmenting}
the control arm with historical data or RWD.


Various approaches have been proposed for dynamic information borrowing, such that the amount of information borrowed from $D_h$ is automatically adjusted according to the congruence between $D_h$ and RCT control data $D_c$. Chen and Ibrahim (Ibrahim and Chen, 2000; Ibrahim \textit{et al}., 2003) proposed a power prior, which controls the degree of information borrowing through a ``power parameter".  Hobbs \textit{et al}. (2011) proposed a commensurate prior that allows for the commensurability of the information in the historical data and current data to determine how much historical information to use. Thall \textit{et al}. (2003), Berry \textit{et al}. (2013), and Chu and Yuan (2018a, 2018b) proposed to use the Bayesian hierarchical model  to borrow information from different data resources or subgroups. Schmidli \textit{et al}. (2014) proposed a robust meta-analytic-predictive (MAP) prior to borrow information from historical data via a mixture prior. 
However, most of these methods have difficulty achieving dynamic information borrowing, leading to substantially inflated type I error and bias,  as noted previously by Neuenschwander \textit{et al}. (2000) and Freidlin and Korn (2013). 

In this paper, we propose a general Bayesian method with elastic priors to address the aforementioned issue. Unlike many existing approaches, the proposed method proactively controls the behavior of dynamic information borrowing through an elastic function, defined as a monotonic function of a congruence measure between $D_h$ and $D_c$. The elastic function is constructed to satisfy a set of prespecified information borrowing constraints.  For example, a borrowing constraint can be set based on a prespecified clinically significant difference such that the amount of borrowing decreases when the difference between $D_h$ and $D_c$ increases. This control leads to a substantially reduced risk of bias. Asymptotically, the elastic prior approach maintains type I and II errors at the nominal value when $D_h$ and $D_c$ are not congruent. 
Our simulation study confirms that, compared to existing methods, the elastic prior approach controls type I errors better, yielding a competitive or higher power. 

Pan \textit{et al}., (2017) proposed the calibrated power prior (CPP) to adaptively borrow information based on the congruence between $D_h$ and $D_c$. There are some important differences between the elastic prior and the CPP. Let $\theta$ denote the parameter of interest (e.g., mean of the efficacy endpoint). The elastic prior is built directly upon the Bayes rule and incorporates $D_h$ through $\theta$'s  posterior $\pi(\theta \, | \, D_h)$ with appropriate inflation of its variance, whereas the CPP takes the power prior framework and works by discounting the likelihood of $D_h$. This makes the elastic prior more general and flexible than the CPP. For example, the elastic prior allows parameter-specific discount by inflating the posterior variance of each parameter separately. In contrast, for the CPP, it is difficult to perform parameter-specific discount because the method discounts the likelihood of $D_h$ as a whole. In addition, the elastic prior can handle multiple $D_h$'s and leverage the existing information-borrowing methodology in a straightforward way. For example, the elastic function can directly be applied to the MAP to adaptively borrow information from multiple $D_h$'s, as described in Section 2.4. Another contribution of this paper is to clarify the intrinsic tradeoff between the power gain and the type I error inflation when borrowing information from $D_h$, and propose a utility-based approach to optimize the elastic prior, which can also be readily used to optimize other methods, including the power prior, commensurate prior, and robust MAP.

The remainder of this article is organized as follows. In Section 2, we introduce the elastic prior method. In Section 3, we evaluate the operating characteristics of the proposed method using simulation, and we conclude with a brief discussion in Section 4.

\section{Methods}
\label{s:method}

Consider a two-arm RCT, let $y$ denote a binary or continuous efficacy endpoint.
Let $\theta_c$ and $\theta_t$ denote $E(y)$ for the control and treatment arms, respectively, and assume that a large value of $\theta_c$ and $\theta_t$ indicates a more desirable outcome. 
The objective of the trial is to compare $\theta_t$ with $\theta_c$ to determine whether the treatment is superior to the control, with the null hypothesis $H_0: \theta_t \le  \theta_c$. Under the Bayesian paradigm, the decision can be made based on the following criterion: reject $H_0$ if $\Pr(\theta_t > \theta_c \, | \, D_c, D_t, D_h)> C$,  where  $C$ is a probability cutoff, and $D_t$ is the treatment data.  We assume that historical data $D_h$ are only available to the control. Thus, we focus on the posterior inference of $\theta_c$ and suppress its subscript when no confusion is caused. In the analysis, the posterior inference for $\theta_t$ will be done using standard Bayesian methods (e.g., using a conventional noninformative or vague prior).

The basic idea of an elastic prior is straightforward. Let $\pi_0(\theta)$ denote a vague initial prior that reflects prior knowledge about $\theta$ before $D_h$ is observed. Applying the prior $\pi_0(\theta)$ to $D_h$, we obtain a posterior distribution $\pi(\theta | D_h)$. The elastic prior is constructed by inflating the variance of $\pi(\theta | D_h)$ by a factor of $g(T)^{-1}$, where $T$ is a congruence measure between $D_h$ and $D_c$, and  $g(T)$ is the elastic function, a monotonic function with values between 0 and 1. When $T\to 0$, reflecting a prefect congruence between $D_h$ and $D_c$, $g(T)\to 1$ and $\pi(\theta | D_h)$  will be used as the prior for full information borrowing. When $T\to \infty$, reflecting substantial incongruence between $D_h$ and $D_c$, $g(T)\to 0$ and the elastic prior will become a noninformative prior.  In the next subsections, we elaborate this approach using binary, Gaussian, and survival endpoints.

\subsection{Binary endpoint}
\label{binary}
\subsubsection{Elastic prior}
Let $n_h$ and $n_c$ respectively denote the sample size of $D_h$ and $D_c$, $D_h=(y_{h,1}, \cdots, y_{h,n_h})$ and $D_c=(y_{c,1}, \cdots, y_{c,n_c})$, where $y_{h,i}\overset{i.i.d}{\sim}Bernoulli(\theta_h)$ and $y_{c,i}\overset{i.i.d}{\sim}Bernoulli(\theta)$.
{{Let}} $\overline{y}_h =  \left. \sum_{i=1}^{n_{h}} y_{h,i} \right/ {n_{h}}$ and $\overline{y}_c = \left. \sum_{i=1}^{n_{c}} y_{c,i} \right/{n_{c}} $.  
Assuming a vague prior $\pi_0(\theta_h)\sim Beta(\alpha_0, \beta_0)$, with  small values of $\alpha_0$ and $\beta_0$ (e.g., $\alpha_0=\beta_0=0.1$), 
given $D_h$, the posterior of $\theta_h$ is given by
\begin{equation*}
\pi(\theta_h | D_{h}) \propto Beta(\alpha_0 + n_h\overline{y}_h, \beta_0 +n_{h}-n_h\overline{y}_h).
\end{equation*}
The elastic prior is defined as
\begin{equation}  \label{binelastic}
\pi^*(\theta | D_{h}) \propto Beta((\alpha_0 + n_h\overline{y}_h)g(T), (\beta_0 + n_{h}-n_h\overline{y}_h)g(T)).
\end{equation}
The elastic prior $\pi^*(\theta | D_{h})$ has the same mean as $\pi(\theta_h | D_{h})$, but 
inflates the latter's variance by a factor of $g(T)^{-1}$.

For a binary endpoint, there are many different choices for congruence measure $T$. For example, we may consider 
$
T={|  \overline{y}_c - \overline{y}_h |}\left/{\sqrt{\overline{y}(1-\overline{y})(\frac{1}{n_c}+\frac{1}{n_h})}}\right. ,
$
where $\overline{y}=({\overline{y}_{c} n_{c}+ \overline{y}_{h} n_{h}})/({n_{c}+n_{h}})$ is a pooled sample mean. While different choices of $T$ have different advantages; in this paper, we choose the Chi-square test statistic:
$$
T=\sum_{j=c, h} \frac{(O_{0 j}-E_{0 j})^2}{E_{0 j}}+\sum_{j=c, h} \frac{(O_{1 j}-E_{1 j})^2}{E_{1 j}},
$$
where $O_{0 j}$ and $O_{1 j}$ are the observed number of responders and non-responders for $D_c$ and $D_h$; $E_{0 j}$ and $E_{1 j}$ are the expected number of responders and non-responders, which are given by
$
E_{0 j}=n_{j} \frac{\sum_{j=c, h} n_{j}-\sum_{j=c, h} \sum_{i=1}^{n_{j}} y_{j,i}}{\sum_{j=c, h} n_{j}}$ and $ E_{1 j}=n_{j} \frac{\sum_{j=c, h} \sum_{i=1}^{n_{j}} y_{j, i}}{\sum_{j=c, h} n_{j}}.
$
A large value of $T \in (0, \infty)$ indicates low congruence between $D_c$ and $D_h$.

Elastic function $g(T)$ serves as a link function that maps congruence measure $T$ to an information discount factor. {Any monotonic function} could be used as an elastic function, as long as $g(T) \to 1$ when the value of $T$ corresponds to congruence and  $g(T) \to 0$ when the value of $T$ corresponds to incongruence. In this paper, we choose a logistic function
\begin{equation} \label{elasticf}
g(T)= \frac{1}{1+\exp \{a+b \times \log (T)\}},
\end{equation}
where $a$ and $b>0$ are prespecified tuning parameters. When appropriate, a more flexible elastic function $
g(T)= \frac{1}{1+\exp [ a+b \times \{ \log (T)\}^c ] }$ can be used to further control the rate of change from borrowing to no borrowing using the additional parameter $c$ (see Figure 1 (a)). 

\begin{figure}[!htbp]
\begin{center}
\includegraphics*[scale=0.7]{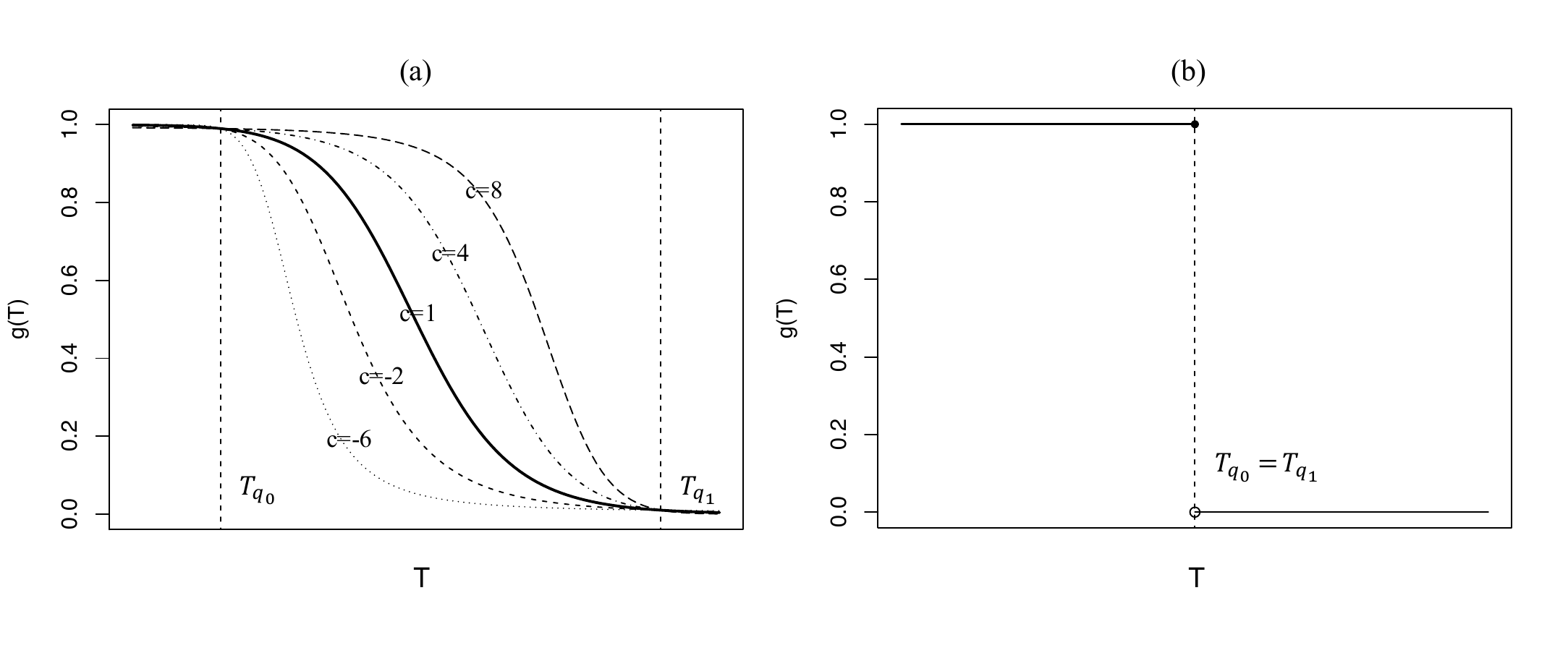}
\end{center}
\caption{(a) { A class of logistic elastic functions defined by  $g(T)= \frac{1}{1+\exp [ a+b \times \{ \log (T)\}^c ] }$, }
and (b) a step elastic function, where $g(T)=1$ leads to full information borrowing and $g(T)=0$ leads to no information borrowing.} \label{fig_elastic}
\end{figure}

\subsubsection{Determination of the elastic function}
This section discusses a procedure to determine the tuning parameters $a$ and $b$ in the elastic function, which is motivated by the following proposition. 

 \noindent{\bf Proposition} $\quad$  For any method that borrows information from historical or other external data, dynamically or non-dynamically, the inflation of type I or II error is inevitable under finite samples, depending on whether historical or other external data under- or over-estimate the treatment effect of the control arm when compared to the current data.

The heuristic justification of the proposition is straightforward: when $\theta_h\neq \theta $, the type I or II error inflates whenever information-borrowing is triggered, depending on $\theta_h < \theta $ or $\theta_h > \theta $. With a finite sample, when  $\theta_h\neq \theta$, there is non-zero probability that the observed $D_h$ and $D_c$ are comparable and thus trigger (inappropriate) information borrowing, thus inflating the type I or II error.  This proposition is important, because it sets a realistic expectation for information-borrowing methods and avoids vain efforts to pursue a dynamic information borrowing method that can strictly control type I errors in finite samples.

\begin{figure}[!htbp]
\begin{center}
\includegraphics*[scale=0.65]{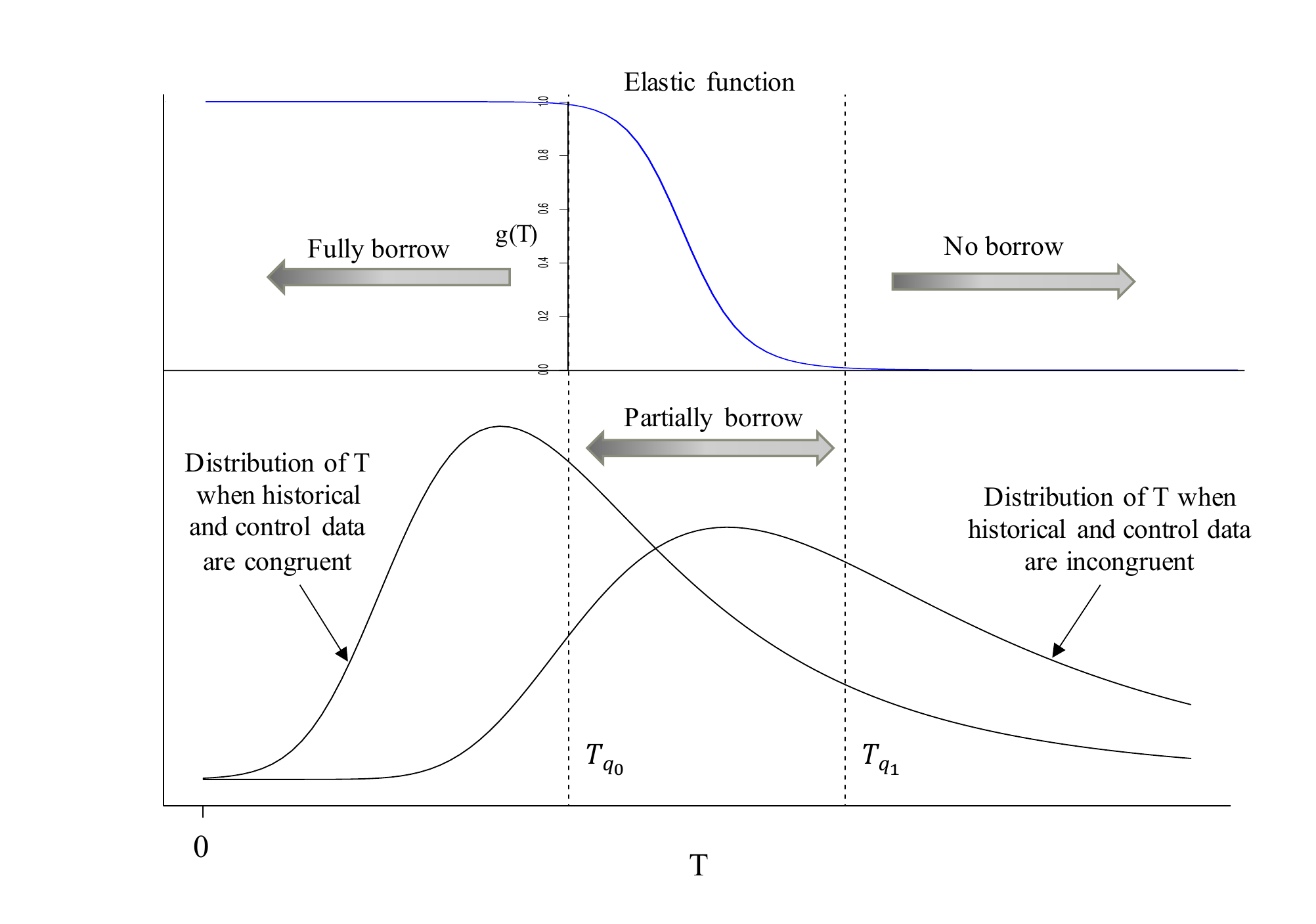}
\end{center}
\caption{Dynamic information borrowing through the elastic function. }\label{T_distri}
\end{figure}

Since the inflation of type I or II errors is inevitable with information borrowing, one reasonable strategy is to control type I and II error inflation according to certain prespecified criteria. This motivates the following procedure to choose the elastic function (\ref{elasticf}), as illustrated in Figure  \ref{T_distri}. Without loss of generality, we assume a larger value of $T$ indicates larger incongruence between $D_h$ and $D_c$.

\begin{enumerate}

\item Elicit from subject matter experts a clinically significant difference between $\theta$ and $\theta_h$ (denoted as $\delta$),  under which $D_h$ and $D_c$ are deemed substantially different such that it is not appropriate to borrow information between $D_h$ and $D_c$.
The determination of $\delta$ often requires communication between sponsors and regulatory bodies. 

\item ({\it Congruent case}) Simulate $R$ replicates of $D_c = (y_{c,1}, \cdots, y_{c,n_c})$ from $Bernoulli (\widehat{\theta}_h)$, with $\widehat{\theta}_h = \overline{y}_h$, and calculate congruence measure $T$ between $D_h$ and each simulated $D_c$, resulting in $\boldsymbol{T}_0 = (T_1, \cdots, T_R)$, where $T_r$ denotes the value of $T$ based on the $r$th simulated $D_c$.

\item ({\it Incongruent cases}) Simulate $R$ replicates of $D_c$ from $Bernoulli (\widehat{\theta}_h +  \delta)$, and calculate congruence measure $T$ between $D_h$ and each simulated $D_c$, resulting in $\boldsymbol{T}_1^+ = (T_1^+, \cdots, T_R^+)$, where $T_r^+$ denotes the value of $T$ based on the $r$th simulated $D_c$. Repeat this with $D_c$ simulated from $Bernoulli (\widehat{\theta}_h -  \delta)$, resulting in $\boldsymbol{T}_1^- = (T_1^-, \cdots, T_R^-)$, where $T_r^-$ denotes the value of $T$ between $D_h$ and the $r$th simulated $D_c$.

\item Let $C_1$ and $C_2$ be constants close to 1 and 0, respectively, e.g., $C_1=0.99$ and $C_2=0.01$, and let $T_{q_0}$ denote the $q_0$th percentile of $\boldsymbol{T}_0$, $T_{q_1}^+$ and $T_{q_1}^-$ denote the $q_1$th percentile of $\boldsymbol{T}_1^+$ and $\boldsymbol{T}_1^-$, respectively, and define $T_{q_1}=min(T_{q_1}^+, T_{q_1}^-)$. Determine the elastic function (\ref{elasticf}) by solving the following two equations:
\begin{equation}
C_1 = g(T_{q_0}) \quad \mathrm{and} \quad C_2 = g(T_{q_1}), \label{twoeqn1}
\end{equation}
where the first equation enforces (approximately) full information borrowing, and the second essentially enforces no information borrowing.
This leads to the solution:
\begin{equation*}  
 a = \log(\frac{1-C_1}{C_1})-\frac{\log(\frac{(1-C_1) C_2}{(1-C_2) C_1})(\log(T_{q_0}))}{\log(T_{q_0})-\log(T_{q_1})} \quad \mathrm{and} \quad  b= \frac{\log (\frac{(1-C_1) C_2}{(1-C_2) C_1})}{\log(T_{q_0})-\log(T_{q_1})}. 
\end{equation*}
\end{enumerate}

Note that in step 4, as incongruence can occur in either direction (i.e., $\theta_c$ is larger or smaller than $\theta_h$), we take $T_{q_1}=min(T_{q_1}^+, T_{q_1}^-)$ to ensure no information borrowing under the more conservative direction. As shown in Figure 2, $q_0$ and $q_1$ define the borrowing and no borrowing regions. Expanding borrowing region increases the power (when $D$ and $D_h$ are truly congruent), at the cost of higher type I error (when $D$ and $D_h$ are truly incongruent); whereas reducing borrowing region provides a better control of type I error, but at the cost of reducing power.  To account for this intrinsic power-type-I-error tradeoff, we propose  
to choose $q_0$ and $q_1$ to maximize the trade-off between the power (in the congruent case) and type I error (in the incongruent case). 

Toward this goal, let $\rho$ denote the power under the congruent case, $\psi$ denote the type I error under the incongruent case described in Step 3, and $\eta$ is a type I error threshold. We define the utility:
\begin{equation}\label{utility}
U(q_0, q_1)=\rho - w_1\psi -w_2(\psi - \eta )I(\psi > \eta),
\end{equation}
where $w_1$ and $w_2$ are penalty weights. $U(q_0, q_1)$ is a function of $q_0$ and $q_1$ because $\rho$ and $\psi$ depend on $g(T)$ (i.e., the elastic prior), and the latter is a function of $q_0$ and $q_1$ as equation (\ref{twoeqn1}). This utility imposes a penalty of $w_1$ for each unit increase of a type I error before it reaches $\eta$, and then a penalty of $w_1+w_2$.  In our simulation, we set $w_1=1$, $w_2=2$, and $\eta=0.1$, which means that before the type I error reaches 0.1, the penalty for a 1\% increase of type I errors is to deduct the power by 1\%; and once the type I error exceeds 0.1, the penalty for a 1\% increase of type I errors increases to deduct the power by 3\%.  Through a grid search (see Appendix A in Supporting Information), we can identify the $(q_0, q_1)$ that maximize $U(q_0, q_1)$. 

A special and limiting form of the logistic elastic function (\ref{elasticf}), obtained with a large value of $b$,  is the following step function
\begin{equation} \label{stepf}
g(T)=\left\{
\begin{array}{rcl}
1 &  & {T \le T_{q_0}}, \\
0 &  & {T > T_{q_0}}.
\end{array}\right.
\end{equation}
As shown in Figure 1 (b), under this step function, $T_{q_0} = T_{q_1}$, and thus  full information borrowing occurs if $T \le T_{q_0}$, and no information borrowing occurs if $T > T_{q_0}$. Compared to logistic elastic function (\ref{elasticf}), one advantage of the step elastic function is that its calibration is simpler, needing only two steps:
\begin{enumerate}
\item[1.] (Congruent case) Simulate $R$ replicates of $D_c = (y_{c,1}, \cdots, y_{c,n_c})$ from $Bernoulli (\widehat{\theta}_h)$, with $\widehat{\theta}_h = \overline{y}_h$, and calculate congruence measure $T$ between $D_h$ and each of the simulated $D_c$'s, resulting in $\boldsymbol{T}_0 = (T_1, \cdots, T_R)$, where $T_r$ denote the value of $T$ based on the $r$th simulated $D_c$.
\item[2.] Use a grid search to identify the $T_{q_0}$ that maximizes utility $U(q_0)$.
\end{enumerate}
Numerical study shows that the step elastic function can achieve similar operating characteristics as a logistic function, but with greater simplicity, making it a good choice for practical use.

\subsubsection{Statistical properties}

The elastic prior has several desirable properties. The proof is provided in the Appendix B in Supporting Information. 

\noindent{\bf Theorem 1} $\quad$  The elastic prior defined in (\ref{binelastic}) is information-borrowing consistent. That is, when $n_h \to \infty $ and $n_c \to \infty$, it achieves full information borrowing if $D_h$ and $D_c$ are congruent (i.e., $\theta_h=\theta$), and discards $D_h$ if $D_h$ and $D_c$ are incongruent (i.e., $\theta_h \neq \theta$).

\noindent{\bf Corollary 1} $\quad$  The elastic prior asymptotically maintains the type I error at the nominal value.

In contrast, most existing dynamic information borrowing methods (e.g.,  the power prior, commensurate prior, and robust MAP) do not have this property. This is because the observation unit used to estimate the information-borrowing parameter (e.g., power parameter, shrinkage parameter) in these approaches is the study, not the subject (Neuenschwander \textit{et al}., 2000; Pan \textit{et al}., 2017). For example, the power parameter and shrinkage parameter represent the between-study variation.  To achieve the information-borrowing consistency, the estimate of the information-borrowing parameter must converge to no borrowing when $D$ and $D_h$ are incongruent, and converge to full borrowing when $D$ and $D_h$ are congruent. This requires the number of historical datasets (not the number of observations within each historical dataset) goes to infinity, which is not the case in practice.

Another desirable property of the elastic prior is that it is straightforward to determine its prior effective sample size (PESS), which is simply $g(T)n_h$, as $g(T)$ is a variance inflation factor.  
In contrast, determining PESS for existing methods (e.g., commensurate prior and robust MAP prior) is more involved, and we found that different PESS calculations used by these methods (Morita \textit{et al}., 2008; Hobbs \textit{et al}., 2013; Neuenschwander \textit{et al}., 2020) often led to substantially different, sometimes improper results (e.g., PESS $> n_h$) (Chen \textit{et al}., 2018; Hobbs \textit{et al}., 2013). 

\subsection{Gaussian and survival endpoints}

Consider a normal endpoint $y_{c,i} \stackrel{iid}{\sim} N(\theta, \sigma^2)$ and $y_{h,i}  \stackrel{iid}{\sim} N(\theta_h, \sigma^2_h)$, with interest in estimating $\theta$. 
With a noninformative prior $\pi_0(\theta_h) \propto 1$ and historical data $D_h$, the posterior of $\theta_h$  is 
$
\pi(\theta_h | D_h, \sigma_h^2) \propto \pi_0(\theta_h)f(D_h | \theta_h, \sigma_h^2) = N( \overline{y}_h,\, \frac{\sigma_h^2}{n_h}).
$
An unknown $\sigma_h^2$ is often replaced by its maximum likelihood estimate $\widehat{\sigma}_h^2 = \left.  \sum_{i=1}^{n_{h}}\left(y_{h, i}-\overline{y}_{h}\right)^{2} \right/ {n_{h}}$.
The elastic prior of $\theta$ is obtained by inflating the variance of $\pi(\theta_h | D_h, \sigma_h^2)$ with the elastic function $g(T)$ as follows:
\begin{equation} \label{normalelastic}
\pi^{*}(\theta | D_{h}, \sigma_{h}^{2}) = N(\overline{y}_{h},\, \frac{\sigma_{h}^{2}}{n_{h}g(T)}).
\end{equation}
Analogue to Section 2.1, the prior effective sample size for $\pi^{*}(\theta | D_{h}, \sigma_{h}^{2}) $ is simply $g(T) n_h$. Full information borrowing is achieved when $g(T)=1$, and no information borrowing occurs when $g(T)=0$. In this scenario, the power prior may obtain similar prior in (\ref{normalelastic}). The key difference is that $g(T)$ is pre-specified to proactively control type I and II error rates and its expected value is known prior to the trial conduct. In addition, as the power prior works by discounting the whole likelihood, it does not allow parameter-specific adaptive information borrowing, for example, when we are interested in estimating and information borrowing on both $\theta$ and $\sigma^2$ as describe later.

The logistic elastic function (\ref{elasticf}) or step elastic function (\ref{stepf}) can be used to dynamically control information borrowing based on the congruence measure $T$. A reasonable choice of $T$ is $t$ statistic
\begin{equation*} 
T= \frac{\left|\overline{y}_{c}-\overline{y}_{h}\right|}{s\sqrt{\frac{1}{n_h} + \frac{1}{n_c}}} ,
\end{equation*}
where $s = \sqrt{\frac{\left(n_{c}-1\right) s_{c}^{2}+\left(n_{h}-1\right) s_{h}^{2}}{n_{c}+n_{h}-2}} $ with $s_{c}^{2}$ and $s_{h}^{2}$ denoting the sample variance of $D_c$ and $D_h$, respectively.  A larger value of $T$  indicates less congruence between  $D_h$ and $D_c$. The choice of $T$ is not unique and can be tailored to quantify inferential interest. For example, if the objective of the trial is to compare the variance between the historical and control arms, the $F$ statistic of testing equal variance is an appropriate measure for the congruence of $D_h$ and $D_c$ in variance. In addition, if subject-level data are available for $D_h$, the Kolmogorov-Smirnov statistic provides a nonparametric congruence measure. The calibration procedure of elastic function $g(T)$ is similar to that for the binary endpoint and provided in the Appendix C in Supporting Information. The construction of the joint elastic prior for $(\theta, \sigma^2)$ is provided in Appendix D in Supporting Information.

Along the same line, the elastic prior method can be applied to a survival endpoint. The logistic elastic function (\ref{elasticf}) or step elastic function (\ref{stepf}) can be used as $g(T)$, where $T$ is the logrank test statistic as a congruence measure between $D_h$ and $D_c$. See Appendix E and F in Supporting Information for details. It can be shown that for both continuous and survival endpoints, the elastic prior is also information-borrowing consistent and asymptotically maintains the type I error at the nominal value, as described in Theorem 1 and Corollary 1.

{\color{black} 

\subsection{Incorporating covariates}
In some applications, a set of $L$ baseline covariates $(x_1, \cdots, x_L)$ are known to be predictive of treatment effects. As the adaptation of the elastic prior relies on the congruence measure $T$,  these covariates can be used, in addition to $y$,  to provide more accurate assessment on the congruence between $D_h$ and $D_c$, thereby achieving more accurate information borrowing. The key question is how to define  $T$ based on $(y, x_1, \cdots, x_L)$? One challenge is that $x$'s and $y$ may be different types of data, e.g., binary, multinomial, continuous or time-to-event. 

We propose to use Fisher's combined p-value method (Fisher, 1934) to synthesize the evidence from $x$'s and $y$ as follows:
 \begin{enumerate}
 \item[1.] Choose an appropriate congruence measure for each of $(y, x_1, \cdots, x_L)$ as described previously, e.g., Chi-square test statistic for categorical variables, $t$ test statistic for continuous variables, and logrank test statistic for survival variables.
 
\item[2.] Convert these $L+1$ test statistics into p values, denoted as $p_0, \cdots, p_L$, where $p_0$ is associated with $y$ and $p_1, \cdots, p_L$ are associated with $x_1, \cdots, x_L$.

\item[3.] Apply Fisher's method to combine them to obtain the congruence measure 
$
T=-\log(p^*),
$
with the combined p value $p^*=\Pr\{\chi^2>-2\sum_{\ell=0}^{L} \log(p_\ell)\}$, where $\chi^2$ is a Chi-square distributed random variable with a degree of freedom of  $2(L+1)$.
\end{enumerate}

In the case that different variables are believed to have different importance to define congruence, a weighted value of combined p value can be used in step 3 to generate $T$, i.e., $p^*=\Pr\{\chi^2>-2\sum_{\ell=0}^{L} w_\ell \log(p_\ell)\}$, where $w_\ell$ is the importance weight. Given the congruence measure $T$, the same approach described previously can be used to define the elastic prior. The calibration of elastic function $g(T)$ is provided in Appendix G in Supporting Information.

\subsection{Extension to multiple historical datasets}
The proposed method can be extended to borrow information from $K$ independent historical datasets $D_{h, 1}, \cdots, D_{h, K}$ as follows:
\begin{enumerate}
\item[1.] Starting with noninformative prior $\pi_0(\theta)$, apply the method described previously to each of $D_{h, 1}, \cdots, D_{h, K}$ to obtain the data-specific elastic prior  $\pi^*(\theta | D_{h, k})$, $k=1, \cdots, K$.

\item[2.] The (overall) elastic prior 
$ \pi^*(\theta | D_{h, 1}, \cdots, D_{h, K}) = \left. \prod_{k=1}^K\pi^*(\theta | D_{h, k}) \right/ \pi_0(\theta)^{K-1}$.
\end{enumerate}
The derivation of this prior is provided in Appendix H in Supporting Information. Because elastic functions $g(T_1), \cdots, g(T_K)$ are determined independently based on $D_{h,1}, \cdots, D_{h, K}$, one advantage of this elastic prior is that its allow study-specific dynamic information borrowing with minimal interference among $D_{h, 1}, \cdots, D_{h, K}$. For example, if $D_{h, 1}$ is congruent to $D_c$ and $D_{h, 2}$ is not congruent to $D_c$, the elastic prior will borrow more information from $D_{h, 1}$ and less information from $D_{h, 2}$. This method is applied to binary, normal and survival endpoints with an appropriate choice of $T$, e.g., Chi-square statistic, $t$ statistic and logrank test statistic for Binary, Gaussian and survival endpoints, respectively.

Another approach is to aggregate historical information through meta-analysis of $D_{h, 1}, \cdots, D_{h, K}$, and then construct the elastic prior. This can be done using two steps: (1) perform meta-analysis on $D_{h, 1}, \cdots, D_{h, K}$ using the Bayesian hierarchical (or random-effects) model to obtain the posterior predictive distribution of $\theta$ (i.e., MAP prior), $\pi(\theta | D_{h, 1}, \cdots, D_{h, K})$; (2) inflate the variance of the MAP prior using the elastic function $g(T)$ to obtain the elastic prior $\pi^*(\theta | D_{h, 1}, \cdots, D_{h, K})$. One challenge is how to choose an appropriate statistic $T$ to measure the congruence between $D_c$ and $K$ datasets.  The congruence measure $T$ discussed previously is applicable to each of $D_{h, 1}, \cdots, D_{h, K}$, but it is not clear how to combine them into a single global congruence measure. To address this issue, we borrow the concept of the posterior predictive model assessment method (Gelman \textit{et al}., 1996; Gelman \textit{et al}., 2013). The basic idea is that if $D_c$ is congruent to $D_{h, 1}, \cdots, D_{h, K}$, we could expect that the actual observed $D_c$ will be generally consistent with the data generated from $\pi(D_c | D_{h, 1}, \cdots, D_{h, K})$. Therefore, if the observed $D_c$ is located on the far tail of the predicted distribution of $\pi(D_c | D_{h, 1}, \cdots, D_{h, K})$, then $D_c$ is likely to be incongruent to the historical datasets.  This motivates us to use the posterior predictive p value as the congruence measure $T$. This approach is general and also can be used for a single historical dataset with various endpoints. Using a normal endpoint as example, $T$ is calculated as follows:

\begin{enumerate}
\item[1.] Draw $R$  samples of $\theta$ from $\pi(\theta|D_{h,1}, \cdots, D_{h, K})$, denoted as $\theta_{(1)}, \cdots, \theta_{(R)}$. Given $\theta_{(r)}$,  simulate trial data $D_c=(y_{c,1}, \cdots, y_{c, n_c})$, and denote its sample mean as $\overline{y}^{(r)}_c$, $r=1, \cdots, R$. In our simulation, we use  $R=10,000$.
\item[2.] Let $\overline{y}_c$ denote the actual observed sample mean of $D_c$; the congruence measure is defined as
$
T=-\log(PP),
$
where $PP=2 \times min(\sum_{r=1}^R I(\overline{y}_c^{(r)}>\overline{y}_c)/R, \sum_{r=1}^R I(\overline{y}_c^{(r)}<\overline{y}_c)/R)$ is the  two-sided posterior-predictive p value.
\end{enumerate}
Of note, the statistical properties of the elastic prior discussed in Section 2.1 (e.g., Theorem 1 and Corollary 1) also apply to those described in  Sections 2.3-2.4.

\section{Simulation studies} \label{s:sim}
In this section, we evaluate the finite-sample properties of the elastic prior approach and compare them to some existing methods.
\subsection{Simulation setting}\label{simuset}
We considered scenarios that involve a two-arm superiority trial with one historical dataset $D_c$, where the endpoint is continuous, binary and time-to-event.
For a continuous endpoint, the sample sizes for historical data, control arm, and treatment arm were $n_h =  50$, $n_c = 25$, and $n_t = 50$, respectively. We generated control data $D_c$ from $N(\theta_c, 1^2)$ with $\theta_c=1$, and treatment data $D_t$  from $N(\theta_t, 1^2)$ with $\theta_t=1$ and $1.5$. We generated the historical data $D_h$ from $N(\theta_h, 1^2)$ and varied its mean $\theta_h$ to simulate the scenarios where $D_h$ is congruent or incongruent to $D_c$.
For a binary endpoint, the sample sizes for the historical data, control arm, and treatment arm were $n_h =  100$, $n_c = 40$, and $n_t = 80$, respectively. We generated $D_c$ from $Bernoulli(\theta_c)$ with $\theta_c=0.4$, and $D_t$ from $Bernoulli(\theta_t)$ with $\theta_t=0.4$, 0.55, and 0.6. We generated $D_h$  from $Bernoulli(\theta_h)$ and varied its mean $\theta_h$ to simulate the scenarios where $D_h$ is congruent or incongruent to $D_c$.  The simulation setting for the survival endpoint is provided in Appendix I in Supporting Information.


We compared the proposed elastic prior with the commensurate prior (CP), (normalized) power prior (PP), and conventional non-informative prior (NP) that ignores historical data. 
For CP, we considered two priors for its shrinkage parameter $\tau$ used in publications:  $\log(\tau)\sim Unif(-30, 30)$ (Hobbs \textit{et al}., 2011) (denoted as CP1), and spike-and-slab prior with a slab of (1, 2), spike of 20 and Pr(slab)=0.98 (denoted as CP2) (Chen \textit{et al}., 2018). For PP, uniform prior $Unif(0, 1)$ is used for the power parameter. For EP1 and EP2, we set $w_1=1$, $w_2=2$, and $\eta=0.1$ in utility, and $\delta=$1 (continuous endpoint) and 0.24 (binary endpoint) to determine the elastic function. For fair comparison, the same criterion is used across the methods to evaluate the efficacy of the treatment, i.e., the treatment is deemed superior to the control if $\Pr( \theta_t -\theta_c>0  \, | \, D_c, D_t, D_h)> C$ for binary and continuous endpoints. 
The probability cutoff $C$ is calibrated for each method with 10,000 simulated trials such that under the null (i.e., corresponding to scenario 1 in Tables  \ref{tab:normal} and \ref{tab:binary}), the type I error is 5\%. The treatment arm does not involve information borrowing and the posterior of $\theta_t$ is obtained based on the conventional noninformative prior. Under other simulation configurations, we conducted 1000 simulations.

\newcommand{\tabincell}[2]{\begin{tabular}{@{}#1@{}}#2\end{tabular}}
\begin{table}[!htp]
\begin{center}
\caption {Simulation results for a normal endpoint using a noninformative prior (NP), elastic prior with the logistic elastic function (EP1) and step elastic function (EP2), commensurate prior with uniform prior (CP1) and spike-and-slab prior (CP2), and power prior (PP). } \label{tab:normal}
\renewcommand\arraystretch{1.75}
\resizebox{\textwidth}{!}{
\begin{tabular}{cccccccccccc}
\toprule
&&&&  \multicolumn{6}{c}{\textbf{Percentage of claiming efficacy (PESS)}} \\ \cline{5-10}
\textbf{Scenario} & $\theta_h$ & $\theta_c$ & $\theta_t$ & \textbf{NP}&   \textbf{EP1} &   \textbf{EP2} & \textbf{CP1} & \textbf{CP2} & \textbf{PP} \\ \hline
 \multicolumn{2}{c}{\textbf{Congruent}} &&&&&&&\\
 1$^*$	&1	&1	&1	&5.0  &5.06(48.9) & 5.17(49.2)	&5.23(42.4) &4.92(4.4)		&5.04(28.7)	\\  
 2	&1	&1	&1.5	&66.3  &93.4(48.9) &93.6(49.2)	&91.6(42.4) &68.4(4.4)		&88.3(28.7)	\\
3	&0.9	&1	&1.5	&66.3  &94.6(48.3)	&95.0(48.8) &93.0(40.7) &69.7(4.4)		&92.0(28.5)	\\
4	&1.1	&1	&1.5	&66.3  &86.0(48.1)	&86.3(48.8) &85.5(40.7) &67.7(4.4)		&83.1(28.5)   \\
\hline
 \multicolumn{2}{c}{\textbf{Incongruent}} &&&&&&\\
5$^*$	&{0}	&{1}	&{1}	 &{5.0}  &{7.7(0.3)} &7.3(0.4)	 &{14.6(0.4)} &{7.0(3.6)}	&{30.0(9.6)}	\\
6$^*$	&{-0.5}	&{1}	&{1}	 &{5.0}  &{7.1(0.0)} &7.0(0.0)	 &{10.0(1.3)} &{8.8(3.2)}	&{18.3(3.2)}	\\
7	&2	&1	&1.5 &66.3 &		72.3(0.2) &72.6(0.4)	&57.6(0.4) &60.7(3.6)		&37.8(9.5)	\\
8	&2.5	&1	&1.5 &66.3 & 	72.3(0.0) &72.6(0.0)	& 69.2(1.4) &57.6(3.3)		&49.1(3.1)	\\
\bottomrule
*Type I error
\end{tabular}}
\end{center}
\end{table}

\begin{table}[!htp]
\begin{center}
\caption {Simulation results for a binary endpoint using a noninformative prior (NP), elastic prior with the logistic elastic function (EP1) and step elastic function (EP2), commensurate prior with uniform prior (CP1) and spike-and-slab prior (CP2), and power prior (PP).} \label{tab:binary}
\renewcommand\arraystretch{1.75}
\resizebox{\textwidth}{!}{
\begin{tabular}{ccccccccccc}
\toprule
&&&&  \multicolumn{6}{c}{\textbf{Percentage of claiming efficacy (PESS)}} \\ \cline{5-10}
\textbf{Scenario} & $\theta_h$ & $\theta_c$ & $\theta_t$ & \textbf{NP}  & \textbf{EP1} & \textbf{EP2} & \textbf{CP1} & \textbf{CP2} & \textbf{PP}\\ \hline
 \multicolumn{2}{c}{\textbf{Congruent}} &&&&&&&\\
1$^*$	&0.4	&0.4	&0.4	&5.09	&5.04(92.4) &4.97(92.4)	&5.10(108.7)	&5.27(5.9)	&5.16(53.8)	\\
2	&0.4	&0.4	&0.6 &75.4	&91.3(92.4)	&90.8(92.4) &91.2(108.7)	&76.0(5.9)	&90.1(53.8)	\\
3	&0.35	&0.4	&0.55	&54.2 &80.6(86.1) &80.9(86.0)	&83.1(103.4)	&58.4(6.2)	&79.9(52.9)	\\
4	&0.42	&0.4	&0.6	&75.4	&88.0(90.1) &87.3(89.7)	&90.0(107.5)	&75.0(5.8)	&87.3(53.6)	\\\hline
 \multicolumn{2}{c}{\textbf{Incongruent}} &&&&&&&\\
5$^*$	&{0.16}	&{0.4}	&{0.4} &{5.0}	&{10.3(6.6)} &6.5(3.10) &{23.7(11.3)}	&{13.9(6.8)}		&{30.1(28.0)}\\
6$^*$	&{0.10}	&{0.4}	&{0.4} &{5.0} &{7.0(1.8)} &6.0(0.20) &{14.8(1.5)}	&{18.6(7.2)}		&{24.6(15.5)}\\
7	&0.6	&0.4	&0.55	&54.2	&48.4(22.0) &46.0(21.2)	&18.6(29.7)	&41.0(4.2)	&22.2(38.8)	\\
8	&0.6	&0.4	&0.6	&75.4	&67.6(22.0)	&65.9(21.2) &38.2(29.7)	&64.8(4.2)	&43.8(38.8)	\\
\bottomrule
*Type I error
\end{tabular}}
\end{center}
\end{table}


\subsection{Simulation results}

Table \ref{tab:normal} shows the results for a normal endpoint. In scenarios 1 and 2, $D_h$ and $D_c$ are congruent. When the treatment is not effective (i.e., scenario 1), all methods control the type I error rate at its nominal value of 5\%. When the treatment is effective (i.e., scenario 2), EP1, EP2, CP1, and PP offer substantial power gain over NP. For example, the power of EP1 is 27.1\% higher than NP, and also slightly higher than CP1 and PP.  EP2 has comparable performance to EP1. In contrast, CP2 provides little power improvement, indicating that the spike-and-slab prior is too conservative to borrow information. Similar results are observed in scenarios 3 to 4, where $D_h$ and $D_c$ are approximately congruent. Scenarios 5-8 consider the case that $D_h$ and $D_c$ are incongruent. Specifically, in scenarios 5 and 6, the treatment is not effective, and the results are type I errors. Compared to CP1 and PP, EP1 and EP2 offer better type I error control. For example, in scenario 5, the type I errors of EP1 and EP2 are 7.7\% and 7.3\%, whereas the type I errors of CP1 and PP are 14.6\% and 30\%, respectively.  CP2 has little type I inflation because it barely borrows information, demonstrated by its low power when the $D_h$ and $D_c$ are congruent (i.e., scenarios 3 and 4). In scenarios 7 and 8, the treatment is effective, and the results are power. EP1 and EP2 yield higher power to detect the treatment effect than CP1 and PP.  For example, in scenario 7, the power of EP2 is 15.0\% and 34.8\% higher than CP1 and PP, respectively.

Table \ref{tab:binary} shows the results for a binary endpoint, which are generally consistent with these for normal endpoint. Scenarios 1 to 4 consider the case that $D_h$ and $D_c$ are congruent or approximately congruent.  In scenario 1, the treatment is not effective; all methods control the type I error rate at its nominal value of 5\%. In scenario 2, the treatment is effective; EP1, EP2, CP1, and PP offer substantial power gain over NP. For example, the power of EP1 is 15.9\% higher than NP, and comparable to CP1 and PP.  Akin to the normal endpoint, CP2 is  similar to NP with little information borrowing. Similar results are observed in scenarios 3 and 4, where $D_h$ and $D_c$ are approximately congruent. Scenarios 5-8 consider the case that $D_h$ and $D_c$ are incongruent. Specifically, in scenarios 5 and 6, the treatment is not effective, and the results are type I errors. { Compared to CP1, CP2 and PP, EP1 and EP2 offer better type I error control. For example, in scenario 5, the type I error of EP1 and EP2 is approximately 1/2  and 1/4 of that of CP1, 1/3  and 1/5 of PP, and 3.6\% and 7.4\% lower than CP2. }
In scenarios 7 and 8, the treatment is effective, and the results are power. EP1 and EP2 yield higher power to detect the treatment effect than found with CP1 and PP.  For example, in scenario 7, the power of EP1 and EP2 are more than double that of CP1 and PP.

The results for a survival endpoint are similar to these for normal and binary endpoints (see Table 1 in Supporting Information). Compared to PP, CP1 and CP2, the elastic prior provides better type I error control, and yields comparable or higher power.


\subsection{Multiple historical datasets}
Taking a similar setting as the simulation with one historical dataset, we generated control arm data $D_c$ from $N(\theta_c, 1^2)$ with $\theta_c=1$ and sample size $n_c=25$, and treatment arm data $D_t$  from $N(\theta_t, 1^2)$ with $\theta_t=1, 1.5$ and sample size $n_t=50$.  We considered four historical datasets with sample size 40, 50, 45, and 55, respectively, generated from the following hierarchical model:
$$
\begin{array}{c}
 y_{k} \sim N(\theta_{k}, 1^2), k=1, \cdots, 4, \\
\theta_{1}, \theta_{2}, \theta_{3}, \theta_{4} \sim N(\theta_h, 0.1^2). \\
\end{array}
$$
We varied $\theta_h$ to simulate scenarios where $D_h$ is congruent or incongruent to $D_c$. Similarly, we considered both the logistic elastic and step elastic functions, and denoted them as elastic MAP 1 (EMAP1) and elastic MAP 2 (EMAP2), respectively. We set $\delta=1$ to calibrate the elastic functions.

We compared the elastic MAP priors with the robust MAP prior. Following Schmidli \textit{et al}. (2014), we considered two versions of the robust MAP prior: Mix50 with a weight of 0.5 and the Mix90 design with a weight of 0.1 assigned to MAP. As the benchmark, we also considered the conventional NP that ignores historical data. The treatment is deemed superior to the control if $\Pr( \theta_t -\theta_c>0  \, | \, D_c, D_t, D_h)> C$.  The probability cutoff $C$ is calibrated for each method with 10,000 simulated trials such that under the null (i.e., $\theta_c=\theta_t=\theta_h$, corresponding to scenario 1 in Table \ref{tab:multiple}), the type I error is 5\%. Under other simulation configurations, we conducted 1000 simulations.

\begin{table}[!htp]
\begin{center}
\caption {Simulation results for multiple historical studies using a noninformative prior (NP), elastic MAP prior with the logistic elastic function (EMAP1) and step elastic function (EMAP2), robust MAP priors with 50\% mixture (Mix50) and 90\% mixture (Mix90).} \label{tab:multiple}
\renewcommand\arraystretch{1.75}
\resizebox{\textwidth}{!}{
\begin{tabular}{ccccccccc}
\toprule
&&&&  \multicolumn{5}{c}{\textbf{Percentage of claiming efficacy (PESS)}} \\ \cline{5-9}
\textbf{Scenario} & $\theta_h$ & $\theta_c$ & $\theta_t$ & \textbf{NP} &  \textbf{EMAP1} &  \textbf{EMAP2} & \textbf{Mix50} & \textbf{Mix90} \\ \hline
 \multicolumn{2}{c}{\textbf{Congruent}} &&&&&\\
 1$^*$	&1	&1	&1 &5.00	&5.18(29.3) &5.37(29.1)	&4.99(15.0)	&4.91(29.2) \\
 2	&1	&1	&1.5	&66.3  &91.0(29.3) &90.7(29.1)	&90.1(15.0)	&90.9(29.2) \\
3	&0.9	&1	&1.5	&66.3  &94.2(29.1) 	&93.3(28.9) &93.1(15.0)	&94.1(29.2) \\
 4	&1.1	&1	&1.5	 &66.3  &84.6(28.9) &84.3(28.6)	&83.9(15.0)	&84.2(29.2)\\
\hline
 \multicolumn{2}{c}{\textbf{Incongruent}} &&&&\\
5$^*$	&{0}	&{1}	&{1}  &{5.0} &{8.5(0.7)}	&7.8(0.4) &{14.1(15.0)}	&{26.4(29.2)} \\
6$^*$	&{-0.2}	&{1}	&{1}  &{5.0} &{7.7(0.0)} &7.6(0.0)	&{9.7(15.0)}	&{16.2(29.2)} \\
7 &1.6	&1	&1.5	 &66.3 &61.3(11.5)	& 67.1(8.4) &43.8(15.0)&35.4(29.2)\\
8	&2	&1	&1.5 &66.3 &75.3(0.0)	&75.1(0.2) &63.0(15.0)	&47.8(29.2) \\
\bottomrule
*Type I error
\end{tabular}}
\end{center}
\end{table}

Table \ref{tab:multiple} shows the results. When historical data and control data are congruent (i.e., scenarios 1 to 4), EMAP1 and EMAP2 have comparable performance to Mix50 and Mix90. All methods control type I errors at the nominal value of 5\% (scenario 1), and they yield substantially higher power than the NB due to borrowing information from historical datasets.  Scenarios 5-8 consider the case that historical data and control data are incongruent. In scenarios 5-6, the treatment is ineffective and the results are type I errors. EMAP1 and EMAP2 offer better type I error control than the robust MAP. For example, in scenario 5, the type I error of EMAP1 and EMAP2 are 8.5\% and 7.8\%, whereas that of Mix50 and Mix90 are 14.1\% and 26.4\%, respectively. In addition, EMAP1 and EMAP2 provide substantial power gain over Mix50 and Mix90. For example, in scenario 7, the power of EMAP1 is 17.5\% and 25.9\% higher than Mix50 and Mix90, respectively, and EMAP2 has 23.3\% and 31.7\% higher power than Mix50 and Mix90, respectively.

\subsection{Incorporating covariates}
Lastly, we briefly evaluated the performance of elastic prior when incorporating covariates. The results (see Appendix J in Supporting Information) show that incorporating covariates improves the performance of the elastic prior with comparable type I error control and higher power.

}

\section{Discussion}
\label{s:conclu}

We have proposed the elastic prior to dynamically borrow information from historical data.  Through the use of elastic function, the elastic prior approach adaptively borrows information based on the congruence between trial data and historical data. The elastic function is constructed based on a set of information-borrowing constraints prespecified such that the prior will borrow information when historical and trial data are congruent, and refrain from information borrowing when historical and trial data are incongruent. The elastic prior is information-borrowing consistent, and is easy to quantify using a prior effective sample size. Simulation study shows that, compared to existing methods, the elastic prior has better type I error control, and yields competitive or higher power. In addition, we provide insights on what can and cannot be achieved using the information-borrowing method, which is useful for guiding future methodology development.

The good performance of the elastic prior stems from the use of elastic function to regulate the behavior of information borrowing within the range of the parameter space of practical interest. That is, the elastic prior does not completely rely on the data to determine information borrowing. It also incorporates the subject matter knowledge (e.g., when it should borrow or not) to enhance and govern the performance. In contrast, many existing methods intend to achieve dynamic information borrowing by estimating the information-borrowing parameter (e.g., power parameter in power prior or shrinkage parameter in commensurate prior), jointly with model parameters, based on data. However, the data contain extremely limited information for estimating the information-borrowing parameter because the observation unit contributing to the estimation is the dataset, not subject-level observations. For example, one historical dataset and one trial dataset actually provide only two observations to estimate the information-borrowing parameter. This is a well-known issue in meta-analysis for estimating the between-study variation. As a result, these dynamic information borrowing methods cannot reliably sense the congruence/incongruence between historical data and trial data to perform appropriate information borrowing.

The idea of an elastic prior is general, and it also can be applied to both commensurate and power priors to improve their operating characteristics. We outline this approach in the Appendix K in Supporting Information. In addition, we have focused on two-arm randomized superiority trials with binary, normal or survival endpoints.  The methodology can be applied to single-arm and multiple-arm trials, as well as other types of trials, for example, noninferiority and  equivalence trials. 

\bigskip
\bigskip
\bigskip
\bigskip
\noindent{ \bf Disclaimer} \\
This article reflects the views of the author, and it should not be construed to represent FDA views or policies.



\clearpage
\begin{center}
\textbf{\large Appendix}
\end{center}

 \noindent\textit{A. Grid search for percentile combination $(q_0, q_1)$} \\
 Let $(q_0^{(1)}, \cdots, q_0^{(J)})$ and $(q_1^{(1)}, \cdots, q_1^{(K)})$ denote the prespecified searching grid for $q_0$ and $q_1$, respectively. We used $q_0^{(1)}=q_1^{(1)}=0.3$ and $q_0^{(J)}=q_1^{(K)}=0.9$, and set a grid step of 0.1. The following steps are used to find the $(q_0, q_1)$ that optimizes the utility $U(q_0, q_1)$.
 \begin{enumerate}
\item  Given a specific grid $(q_0^{(j)}, q_1^{(k)})$, determine the elastic function using equation (3) in Section 2.
\item Given the obtained elastic function, under the congruent case ($\theta_h=\theta_c$), calibrate the probability cutoff $C$ to control the type I error rate at a nominal value of 5\% and compute the power ($\rho$) through simulation.
\item Given the cutoff $C$, compute the type I error ($\psi$) under the incongruent case (e.g., $\theta_c=\theta_h- \delta$).
\item Identify $(q_0^{(j)}, q_1^{(k)})$ that produces the largest value of $U(q_0^{(j)}, q_1^{(k)})  = \rho - w_1\psi -w_2(\psi - \eta )I(\psi > \eta)$.
\end{enumerate}

For the step elastic function, the calibration of $q_0$ is similar to that shown above. The main difference is that we only need to search over a one-dimensional grid $(q_0^{(1)}, \cdots, q_0^{(J)})$, which greatly reduces the optimization time.

\bigskip\bigskip\bigskip

\noindent\textit{B. Proof of Theorem 1 and Corollary 1} \\
\noindent\textit{B1. Theorem 1} \\
{\color{black} Suppose $T$ is consistent for testing the congruence between the distributions of historical data $D_h$ and control data $D_c$, i.e., $T\to0$ for $n_h, n_c\to + \infty$, when the distribution of $D_h$ is identical to that of $D_c$ (i.e., $D_h$ and $D_c$ are congruent), $T\to + \infty$  otherwise. Consequently, when $D_h$ and $D_c$ are congruent, $T\to0$ for $n_h, n_c\to + \infty$, and thus $g(T)= \frac{1}{1+\exp \{a+b \times \log (T)\}} \to 1$, given $b>0$. The elastic prior fully borrows historical information.  When $D_h$ and $D_c$ are incongruent, $T\to+\infty$ for $n_h, n_c\to +\infty$, and thus $g(T)= \frac{1}{1+\exp \{a+b \times \log (T)\}} \to 0$. No historical information will be borrowed. This theorem holds for any statistics $T$ which satisfies testing consistency, including all the statistics described in Section 2.}

\noindent\textit{B2. Corollary 1} \\
{\color{black} When $D_h$ and $D_c$ are incongruent, $g(T)^{-1} \to +\infty$ due to $T\to +\infty$ for $n_h, n_c\to + \infty$, and thus elastic prior avoids borrowing information from $D_h$ by largely inflating the variance of $\pi(\theta|D_h)$ with a factor $g(T)^{-1}$. This leads to a reduced risk of bias which favors treatment (i.e., $\theta_h < \theta$). Consequently, the type I error is asymptotically maintained at a nominal value.}

\bigskip\bigskip\bigskip
\noindent\textit{C. Determination of elastic function for a normal endpoint} \\
The steps to determine elastic function are similar to those for the binary endpoint, and described as follows:
\begin{enumerate}
\item Estimate the mean and variance of $D_h$ by $\widehat{\theta}_h= \overline{y}_h$ and $\widehat{\sigma}_h^2 = \sum_{i=1}^{n_h} (y_{h,i}-\overline{y}_h)^2/(n_h-1)$ with $\overline{y}_h = \sum_{i=1}^{n_h} y_{h,i} /n_h$.

\item Elicit from subject matter experts a clinically significant difference $\delta$ between $\theta$ and $\theta_h$.

\item ({\it Congruent case}) Simulate $R$ replicates of $D_c = (y_{c,1}, \cdots, y_{c,n_c})$ from $N(\widehat{\theta}_h, \widehat{\sigma}_h^2)$, and calculate congruence measure $T$ between $D_h$ and each simulated $D_c$, resulting in $\boldsymbol{T}_0 = (T_1, \cdots, T_R)$, where $T_r$ denote the value of $T$ based on the $r$th simulated $D_c$.

\item ({\it Incongruent cases}) Simulate $R$ replicates of $D_c$ from $N(\widehat{\theta}_h + \delta, \widehat{\sigma}_h^2)$, and calculate congruence measure $T$ between $D_h$ and each simulated $D_c$, resulting in $\boldsymbol{T}_1^+ = (T_1^+, \cdots, T_R^+)$, where $T_r^+$ denote the value of $T$ based on the $r$th simulated $D_c$. Repeat this with $D_c$ simulated from $N(\widehat{\theta}_h - \delta, \widehat{\sigma}_h^2)$, resulting in $\boldsymbol{T}_1^- = (T_1^-, \cdots, T_R^-)$, where $T_r^-$ denote the value of $T$ between $D_h$ and the $r$th simulated $D_c$.

\item Let $C_1$ and $C_2$ be constants close to 1 and 0, respectively, e.g., $C_1=0.99$ and $C_2=0.01$, and let $T_{q_0}$ denotes the $q_0$th percentile of $\boldsymbol{T}_0$, $T_{q_1}^+$ and $T_{q_1}^-$ denote the $q_1$th percentile of $\boldsymbol{T}_1^+$ and $\boldsymbol{T}_1^-$, respectively, and define $T_{q_1}=min(T_{q_1}^+, T_{q_1}^-)$.

\item Based on $T_{q_0}$ and $T_{q_1}$, determine the elastic function (2) by equation (3) in Section 2.
\end{enumerate}

\bigskip\bigskip\bigskip
\noindent\textit{D. Joint elastic prior of $(\theta, \sigma^2)$ for a normal endpoint} \\
If estimation of $\theta$ and $\sigma^2$ is of interest, we can also construct the joint elastic prior for $(\theta, \sigma^2)$. 
We first apply the noninformative prior $\pi_{0}(\theta_h, \sigma^{2}_h) \propto({1}/{\sigma^{2}_h})^{m}$ to $D_h$, where $m$ is a constant, resulting in the following posterior,
\begin{equation*}
\begin{aligned} \pi(\theta_h, \sigma^{2}_h | D_{h}) & \propto \pi_{0}(\theta_h, \sigma^{2}_h) f(D_{h} | \theta_h, \sigma^{2}_h) \\
& \propto N_\theta ( \overline{y}_{h}, \frac{\sigma^{2}_h}{n_{h}}) IG_{\sigma_h^2} ( \mu_{h}, \epsilon_{h}^{2}) , \end{aligned}
\end{equation*}
where $IG(\cdot)$ is an inverse gamma distribution with mean $\mu_{h}=\frac{n_{h} \widehat{\sigma}_{h}^{2}}{n_{h}-5+2m}$ and variance $
{\epsilon _h^{2}} = \frac{( n_h\widehat \sigma _h^2 )^2}{( n_h-5 + 2m )^2( \frac{n_h-7}{2} + m)}$.  The joint elastic prior for $(\theta, \sigma^2)$ is obtained by inflating the variance of $\pi(\theta_h, \sigma^{2}_h | D_{h}) $ with two elastic functions $g_1(T_1)$ and $g_2(T_2)$,
\begin{equation*}
\pi^{*}(\theta, \sigma^{2} | D_{h}) \propto N_\theta(\overline{y}_{h}, \frac{\sigma^{2}}{n_{h} g_{1}(T_1)}) IG_{\sigma^{2}}(\mu_{h}, \frac{\epsilon_{h}^{2}}{g_{2}(T_2)}),
\end{equation*}
where $T_1$ is $t$ statistic or Kolmogorov-Smirnov statistic described in Section 2.2, and $T_2$ is the $F$ statistic of testing equal variance. Allowing parameter-specific information borrowing renders the elastic prior more flexibility than the power prior. Given the elastic prior and trial data $D_c$, the posterior distribution for $(\theta, \sigma^2)$ is
\begin{equation*}
\pi( \theta ,\sigma ^2|D_c,{D_h} ) \propto N_ \theta(\frac{{n_h}{g_1}(T_1){{\overline y }_h} + n_c{\overline y}_c }{{n_h}{g_1}( T_1 ) + n_c},\frac{\sigma ^2}{{n_h}{g_1}(T_1) + n_c}) IG_{\sigma^2}( \alpha ^ * ,\beta ^ * ),
\end{equation*}
where ${\alpha ^ * } = \frac{n_c+4}{2} + (\frac{n_h-7}{2} + m){g_2}(T_2)$, and  ${\beta^*} = \frac{\sum\nolimits_{i = 1}^{n_c} {y_{c,i}^2 + {n_h}{g_1}(T_1)\overline y _h^2}}{2} - \frac{( {n_h}{g_1}(T_1){{\overline y }_h} + n_c{\overline y}_c )^2}{2{n_h}{g_1}( T_1 ) + 2n_c} + \frac{{n_h}\widehat \sigma _h^2}{{n_h} -5 + 2m}[1 + (\frac{{n_h}-7}{2} + m){g_2}( T_2 )]$.

\bigskip\bigskip\bigskip
\noindent\textit{E. Elastic prior for a survival endpoint} \\
We assume that the survival endpoint $t_i$ follows a proportional hazard model
$$
\lambda(t_i|S_i)=\theta(t_i) \exp(\beta S_i),
$$
where $\theta(t_i)$ is the hazard of the control arm, $S_i$ is the group indicator with $S_i=0$ and 1 indicating the control  and treatment arms, respectively,  and $\exp(\beta)$ represents the hazard ratio between the treatment and control arms. We reject the null hypothesis $H_0: \exp(\beta) \le 1$ and conclude that the treatment arm is superior to the control if $\Pr(\exp(\beta)>1|D_h, D_c, D_t) > C$, where $C$ is a probability cutoff calibrated to control the type I error.

We model $\theta(t_i)$ using a flexible piecewise exponential model by partitioning the time $[0, t_{max}]$ into $J$ disjoint intervals $[0, d_1), [d_1, d_2), \cdots, [d_{J-1}, d_J=t_{max}]$, where $t_{max}$ is the maximum observation time of the trial and historical data. We assume a constant hazard $\theta_{j}$ in the $j$th interval $[d_{j-1}, d_j)$.  For the historical data $D_h$, define $y_{h, i}=  min (t_{i}, r_{i})$ as the observed time, where $r_i$ is the independent censoring time, and $\delta_{h, i} = I (y_{h, i}=t_{i})$ as the censoring indicator. 
Assuming an independent vague gamma prior $\theta_{j} \sim Ga(\kappa_j, \upsilon_j)$ with small values for the shape parameter $\kappa_j$ and the rate parameter $\upsilon_j$ (e.g., $\kappa_j=\upsilon_j=0.1$), the posterior of $\theta_{j}$ based on $D_h$ is given by
\begin{equation*}
\pi(\theta_{j}|~D_h)=Ga(\kappa_j+\sum_{i=1}^{n_h}\delta_{h, i,j}, \upsilon_j + \sum_{i=1}^{n_h} e_{h, i,j}),  \quad  j=1, \cdots, J,
\end{equation*}
where $\delta_{h,i, j}=1$ if the $i$th individual from historical arm has a response in the $j$th interval; and $\delta_{h, i, j}=0$ otherwise. $e_{h,i,j}=d_j-d_{j-1}$ if $y_{h,i}>d_j$; $e_{h, i,j}=y_{h,i}-d_{j-1}$ if $y_{h,i} \in [d_{j-1}, d_j)$; and $e_{h,i,j}=0$ otherwise.

The elastic prior of $\theta_{j}, j=1, \cdots, J, $ is obtained by inflating the variance of $\pi(\theta_{j}|~D_h)$ with the elastic function $g(T)$ as follows
\begin{equation} \tag{1}
\pi^*(\theta_{j}|~D_h)=Ga((\kappa_j+\sum_{i=1}^{n_h}\delta_{h, i,j})g(T), (\upsilon_j + \sum_{i=1}^{n_h} e_{h, i,j})g(T)). \label{TTEprior}
\end{equation}
The logistic elastic function (2) and step elastic function (5) can be used as $g(T)$. Full information borrowing is achieved when $g(T)=1$, and no information borrowing occurs when $g(T)=0$.  We use the logrank test statistic $T$ as the congruence measure of $D_h$ and $D_c$, leading to a small value of $T$ (i.e., $g(T)$ close to 1) when $D_h$ and $D_c$ have similar hazard functions (i.e., congruent), and a large value of $T$ (i.e., $g(T)$ close to 0) when $D_h$ and $D_c$ have different hazard functions (i.e., incongruent). 

Similar to binary and Gaussian endpoints, it can be shown that the resulting elastic prior (\ref{TTEprior}) is information-borrowing consistent, as described in Theorem 1.  The calibration of elastic function $g(T)$ is similar to that for binary endpoint and provided in Appendix F in Supporting Information.

\bigskip\bigskip\bigskip

{\color{black}
\noindent\textit{F. Determination of elastic function for a survival endpoint} \\
The steps to determine elastic function are similar to those for the binary and normal endpoints, and described as follows:
\begin{enumerate}

\item[1.] Estimate the hazard $\boldsymbol{\theta_{h}}=(\theta_{h,1}, \cdots, \theta_{h,J})$ of $D_h$ by maximum likelihood estimation $\boldsymbol{\widehat{\theta}_{h}}$ with $\widehat{\theta}_{h,j}=\sum_{i=1}^{n_h}\delta_{h,i,j}/\sum_{i=1}^{n_h}e_{h,i,j}$.

\item[2.] Elicit from subject matter experts a clinically significant difference $\delta$ on the hazard ratio $\theta_{c,j}/\theta_{h,j}$.

\item[3.] ({\it Congruent case}) Simulate $R$ replicates of $D_c = (y_{c,1}, \cdots, y_{c,n_c})$ from piecewise exponential model with hazard $\boldsymbol{\theta_c} = \boldsymbol{\widehat{\theta}_h}$, and calculate congruence measure $T$ between $D_h$ and each simulated $D_c$, resulting in $\boldsymbol{T}_0 = (T_1, \cdots, T_R)$, where $T_r$ denotes the value of $T$ based on the $r$th simulated $D_c$.

\item[4.] ({\it Incongruent cases}) Simulate $R$ replicates of $D_c$ from piecewise exponential model with hazard $\boldsymbol{\theta_c} =\boldsymbol{\widehat{\theta}_h} \delta $, and calculate congruence measure $T$ between $D_h$ and each simulated $D_c$, resulting in $\boldsymbol{T}_1^+ = (T_1^+, \cdots, T_R^+)$, where $T_r^+$ denotes the value of $T$ based on the $r$th simulated $D_c$. Repeat this with $D_c$ simulated from piecewise exponential model with hazard $\boldsymbol{\theta_c} =\boldsymbol{\widehat{\theta}_h}/\delta$, resulting in $\boldsymbol{T}_1^- = (T_1^-, \cdots, T_R^-)$, where $T_r^-$ denotes the value of $T$ between $D_h$ and the $r$th simulated $D_c$.

\item[5.] Let $C_1$ and $C_2$ be constants close to 1 and 0, respectively, e.g., $C_1=0.99$ and $C_2=0.01$, and let $T_{q_0}$ denote the $q_0$th percentile of $\boldsymbol{T}_0$, $T_{q_1}^+$ and $T_{q_1}^-$ denote the $q_1$th percentile of $\boldsymbol{T}_1^+$ and $\boldsymbol{T}_1^-$, respectively, and define $T_{q_1}=min(T_{q_1}^+, T_{q_1}^-)$. 

\item[6.] Based on $T_{q_0}$ and $T_{q_1}$, determine the elastic function (2) by equation (3) in Section 2.

\end{enumerate}

\bigskip\bigskip\bigskip
\noindent\textit{G. Determination of elastic function when incorporating covariates} \\
Denote $(y_0, x_1, \cdots, x_L)$ as $\boldsymbol{z}=(z_0, z_1, \cdots, z_L)$, where outcome $y_0=z_0$ and covariates $(x_1, \cdots, x_L)=(z_1, \cdots, z_L)$. The steps to determine elastic function when incorporating covariates are described as follows:
\begin{enumerate}

\item[1.] Given historical data $D_h$, estimate the interested parameters $\boldsymbol{\theta_{h,\ell}}$ in the distribution $f(\boldsymbol{\theta_{h, \ell}})$ for $z_{h, \ell}$, $\ell=0, \cdots, L$. For example,  if $z_{h, \ell}$ is normal distributed, estimate its mean and variance $\boldsymbol{\theta_{h, \ell}}=(\mu_{h, \ell}, \sigma_{h, \ell}^2)$ by $\widehat{\mu}_{h, \ell}=\overline{z}_{h, \ell}$ and $\widehat{\sigma}_{h, \ell}^2 = \sum_{i=1}^{n_h} (z_{h,\ell,i}-\overline{z}_{h, \ell})^2/(n_h-1)$; if $z_{h, \ell}$ is binary, estimate the mean $\boldsymbol{\theta_{h, \ell}}=\mu_{h, \ell}$ by $\widehat{\mu}_{h, \ell}=\overline{z}_{h, \ell}$.

\item[2.] Elicit from subject matter experts clinically significant differences between each $z_{h,\ell}$ and $z_{c,\ell}$ (denoted as $\delta_\ell$).

\item[3.] ({\it Congruent case}) Simulate $R$ replicates of $D_c = (z_{c,0}, \cdots, z_{c, L})$ by simulating $z_{c, \ell}=(z_{c, \ell, 1}, \cdots, z_{c, \ell, n_c})$ from distribution $f(\boldsymbol{\widehat{\theta}_{h, \ell}})$, $\ell=0,\cdots, L$. For example, if $z_{c,\ell}$ is normal distributed, generate $z_{c, \ell}$ from $f(\boldsymbol{\widehat{\theta}_{h, \ell}})=N(\widehat{\mu}_{h, \ell}, \widehat{\sigma}_{h, \ell}^2)$; if $z_{c, \ell}$ is binary, generate $z_{c, \ell}$ from $f(\boldsymbol{\widehat{\theta}_{h, \ell}})=Bernoulli(\widehat{\mu}_{h, \ell})$. And calculate congruence measure $T$ between $D_h$ and each simulated $D_c$, resulting in $\boldsymbol{T}_0 = (T_1, \cdots, T_R)$, where $T_r$ denotes the value of $T$ based on the $r$th simulated $D_c$.

\item[4.] ({\it Incongruent cases}) Simulate $R$ replicates of $D_c$ by simulating $z_{c, \ell}$ from distribution $f(\boldsymbol{\widehat{\theta}_{h, \ell}}+\delta_ \ell)$, $\ell=0,\cdots, L$. For example, if $z_{c,\ell}$ is normal distributed, generate $z_{c,\ell}$ from $f(\boldsymbol{\widehat{\theta}_{h, \ell}}+\delta_\ell)=N(\widehat{\mu}_{h, \ell}+\delta_\ell, \widehat{\sigma}_{h, \ell}^2)$; if $z_{c, \ell}$ is binary, generate $z_{c, \ell}$ from $f(\boldsymbol{\widehat{\theta}_{h, \ell}}+\delta_\ell)=Bernoulli(\widehat{\mu}_{h, \ell}+\delta_\ell)$. And calculate congruence measure $T$ between $D_h$ and each simulated $D_c$, resulting in $\boldsymbol{T}_1^+ = (T_1^+, \cdots, T_R^+)$, where $T_r^+$ denotes the value of $T$ based on the $r$th simulated $D_c$. Repeat this with $D_c$ simulated from $f(\boldsymbol{\widehat{\theta}_{h, \ell}}-\delta_\ell)$, $\ell=0, \cdots, L$, resulting in $\boldsymbol{T}_1^- = (T_1^-, \cdots, T_R^-)$, where $T_r^-$ denotes the value of $T$ between $D_h$ and the $r$th simulated $D_c$.

\item[5.] Let $C_1$ and $C_2$ be constants close to 1 and 0, respectively, e.g., $C_1=0.99$ and $C_2=0.01$, and let $T_{q_0}$ denote the $q_0$th percentile of $\boldsymbol{T}_0$, $T_{q_1}^+$ and $T_{q_1}^-$ denote the $q_1$th percentile of $\boldsymbol{T}_1^+$ and $\boldsymbol{T}_1^-$, respectively, and define $T_{q_1}=min(T_{q_1}^+, T_{q_1}^-)$. 

\item[6.] Based on $T_{q_0}$ and $T_{q_1}$, determine the elastic function (2) by equation (3) in Section 2.

\end{enumerate}
}

\bigskip\bigskip\bigskip
\noindent\textit{H. Derivation of elastic prior for multiple historical datasets} \\
{\color{black} Given historical datasets $D_1, \cdots, D_K$ and initial prior $\pi_0(\theta)$, the posterior distribution of $\theta$ is given by
\begin{equation*}
\begin{aligned}
\pi(\theta|D_1, \cdots, D_K) & \propto f(D_1, \cdots, D_K|\theta)\pi_0(\theta)\\
&\propto \{\prod_{k=1}^{K} f(D_k|\theta)\}\pi_0(\theta)\\
& \propto \frac{\{\prod_{k=1}^{K} f(D_k|\theta)\}\pi_0(\theta)^{K}}{\pi_0(\theta)^{K-1}}\\
& \propto \frac{\prod_{k=1}^{K} f(D_k|\theta)\pi_0(\theta)}{\pi_0(\theta)^{K-1}}\\
& \propto \frac{\prod_{k=1}^{K} \pi(\theta|D_k)}{\pi_0(\theta)^{K-1}}.
\end{aligned}
\end{equation*}
The elastic prior of $\theta$ is given by 
$$
\pi^*(\theta|D_1, \cdots, D_K) \propto \frac{\prod_{k=1}^{K} \pi^*(\theta|D_k)}{\pi_0(\theta)^{K-1}},
$$
where $\pi^*(\theta|D_k)$ is the elastic prior which is obtained by applying the method described in Section 2 to dataset $D_k$. Taking a binary endpoint as an example, the elastic prior $\pi^*(\theta|D_k)$ discounting the information from $D_k$ is given by $Beta(\alpha_k, \beta_k)$ with $\alpha_k=(\alpha_0+n_k\overline{y}_k)g(T_k)$ and $\beta_k=(\beta_0 + n_k-n_k\overline{y}_k)g(T_k)$. The elastic prior discounting the information from all datasets is given by 
$$
\pi^*(\theta|D_1, \cdots, D_K) \propto \frac{\prod_{k=1}^{K} Beta(\alpha_k, \beta_k)}{Beta(\alpha_0, \beta_0)^{K-1}}.
$$
 }

 \bigskip \bigskip \bigskip
\noindent\textit{I. Simulation setting for a survival endpoint} \\
For a survival endpoint,  the sample sizes for historical data, control arm, and treatment arm were $n_h =  50$, $n_c = 25$, and $n_t = 50$, respectively. Three intervals are used for the piecewise exponential model, partitioned by $d_1=16$ and $d_2=28$. We generated $D_c$ from the piecewise exponential model with hazard $(\theta_1, \theta_2, \theta_3) = (0.01, 0.03, 0.02)$, and $D_t$ with hazard $(\phi_t\theta_1, \phi_t\theta_2, \phi_t\theta_3)$, where $\phi_t=1$ and 1.65. We generated $D_h$ from piecewise exponential model with hazard $(\phi_h\theta_1, \phi_h\theta_2, \phi_h\theta_3)$, where $\phi_h$ was varied to simulate the scenarios where $D_h$ and $D_c$ are congruent or incongruent. We considered the logistic elastic function (2) and step elastic function (5) and denoted them as elastic prior 1 (EP1) and elastic prior 2 (EP2), respectively.

For EP1 and EP2, we set $w_1=1$, $w_2=2$, and $\eta=0.1$ in utility, and $\delta=\exp(1)$ to determine the elastic function. For fair comparison, the same criterion is used across the methods to evaluate the efficacy of the treatment, i.e., the treatment is deemed superior to the control if  $\Pr( \exp(\beta) >1  \, | \, D_c, D_t, D_h)> C$. 
The probability cutoff $C$ is calibrated for each method with 10,000 simulated trials such that under the null (i.e., corresponding to scenario 1 in Tables  \ref{tab:survival}), the type I error is 5\%. Under other simulation configurations, we conducted 1000 simulations.

Table \ref{tab:survival} shows the results for a survival endpoint, which are similar to these for normal and binary endpoints. When $D_h$ and $D_c$ are congruent or approximately congruent (i.e., scenarios 2-4), EP1, EP2, CP1 and PP have higher power than NP, whereas CP2 provides little power gain.  Compared to CP1 and PP,  the proposed EP1 and EP2 yield higher power. For example, in scenario 2, the power of EP1 is 28.9\%, 2.9\%, 20.1\% and 4.7\% higher than NP, CP1, CP2, and PP, respectively, and EP2 has comparable power as EP1. When $D_c$ is incongruent to $D_h$ (i.e., scenarios 5-8), CP1, CP2 and PP have substantially inflated type I errors or reduced power. In contrast, EP1 and EP2 have better controlled type I errors and maintain power. For example, in scenario 5, the type I error rates of CP1, CP2, and PP are inflated to 28.9\%, 16.0\% and 53.2\%, respectively, while EP1 and EP2 control type I error under 10\%. In scenario 7, the power of EP1 is 30.9\%, 22.9\% and 30.4\% higher than CP1, CP2, and PP.

\begin{table}[!htp]
\begin{center}
\caption {Simulation results for a survival endpoint using a noninformative prior (NP), elastic prior with the logistic elastic function (EP1) and step elastic function (EP2), commensurate prior with uniform prior (CP1) and spike-and-slab prior (CP2), and power prior (PP).} \label{tab:survival}
\renewcommand\arraystretch{1.75}
{
\begin{tabular}{cccccccccc}
\hline
&&&  \multicolumn{6}{c}{\textbf{Percentage of claiming efficacy}} \\ \cline{4-9}
\textbf{Scenario} & $\phi_h$ & $\phi_t$ & \textbf{NP}  & \textbf{EP1} & \textbf{EP2} & \textbf{CP1} & \textbf{CP2} & \textbf{PP}\\ \hline
 \multicolumn{2}{c}{\textbf{Congruent}} &&&&&&\\
1$^*$	&1	&1	&5.05	&5.13 	&5.15 &5.14  &5.19	&5.00\\
2	&1	&1.65	&63.8	&92.7	&92.9	&89.8     &72.6	&88.0\\
3	&0.90	&1.65	&63.8	&95.7	&96.2	&95.2     &75.2	 &92.4\\
4	&1.11	&1.65	&63.8	&86.1	&86.2	&82.0     &70.5	&81.8\\\hline
 \multicolumn{2}{c}{\textbf{Incongruent}} &&&&&&\\
5$^*$	&0.37	&1	&5.0	&9.0	&9.0	&28.9     &16.0     &53.2\\
6$^*$	&0.30	&1	&5.0	&8.8	&8.8	&20.4     &19.0	&48.3\\
7	&3.00	&1.65	&63.8	&72.0	&72.3	&41.1    &49.1	&41.6\\
8	&4.06	&1.65	&63.8	&72.0	&72.2	&53.5    &44.1	&48.4\\
\hline
*Type I error
\end{tabular}}
\end{center}
\end{table}

\clearpage

\bigskip \bigskip \bigskip
\noindent\textit{J. Simulation with covariates} \\
We briefly evaluated the performance of elastic prior when incorporating covariates. We considered a continuous endpoint $y$, with one binary covariate $x_1$ and one continuous covariate $x_2$. For $D_c$, we generated $y$ from $N(1, 1.1)$, $x_{1}$ from $Bernoulli(0.4)$ and $x_{2}$ from $N(4, 1)$ with sample size $n_c=25$; and for $D_h$, we generated $y$ from $N(\theta_h, 1.1)$, $x_1$ from $Bernoulli(\theta_{h,1})$ and $x_{2}$ from $N(\theta_{h,2}, 1)$ with sample size $n_h=50$, where $\theta_h$, $\theta_{h,1}$ and $\theta_{h,2}$ are varied to simulate the scenarios where $D_h$ is congruent or incongruent to $D_c$; for $D_t$, we generated $y$ from $N(\theta_t, 1.1)$ with $\theta_t=1, 1.5$ and sample size $n_t=50$. We denote the elastic prior incorporating covariates as EPC1 for logistic elastic function (2) and EPC2 for step elastic function (5), respectively. We used $\delta =  1$, 0.2 and 1.5 for outcome $y$, covariates $x_1$ and $x_2$ to calibrate the elastic function. We compared EPC1 and EPC2 to EP1 and EP2 that ignore covariates.  

Table \ref{tab:covariate} shows the results, suggesting that incorporating covariates improves the performance of the elastic prior. Specifically, when $D_c$ is congruent or approximately congruent to $D_h$ (i.e., scenarios 1-4), EPC1 and EPC2 yield higher power than EP1 and EP2. For example, in scenario 3, the power of EPC1 and EPC2 are 5.6\% and 2\% higher than EP1 and EP2, respectively. When $D_c$ is incongruent to $D_h$ (i.e., scenarios 5-8), EPC1 and EPC2 control the type I error comparably to EP1 and EP2, and yield higher power. For example, in scenario 5, the type I error rates of EP1, EP2, EPC1, and EPC2 are controlled under 10\%, and in scenario 8, the power of EPC1 and EPC2 are 3.1\% and 1.5\% higher than EP1 and EP2, respectively.

\begin{table}[!htp]
\begin{center}
\caption {Simulation results for a normal endpoint without covariates using elastic prior with the logistic elastic function (EP1) and step elastic function (EP2), and with covariates using elastic prior with the logistic elastic function (EPC1) and step elastic function (EPC2). } \label{tab:covariate}
\renewcommand\arraystretch{1.75}
\footnotesize
{
\begin{tabular}{cccccccccccc}
\hline
& \multicolumn{3}{c}{\textbf{Outcome}}& \multicolumn{4}{c}{\textbf{Covariate}} &  \multicolumn{4}{c}{\textbf{Percentage of claiming efficacy (PESS)}} \\ \cline{2-12}
\textbf{Scenario} & $\theta_h$ &$\theta_c$& $\theta_t$ & $\theta_{h,1} $ & $\theta_{h,2}$ & $\theta_{c, 1}$ &$\theta_{c,2}$ & \textbf{EP1} & \textbf{EP2} &  \textbf{EPC1} & \textbf{EPC2}\\ \hline
 \multicolumn{2}{c}{\textbf{Congruent}} &&&&&& &&&&\\
 1$^*$	&1	&1	&1 &0.4	&4	&0.4	&4 &5.10(44.6)	&5.17(48.7)	&5.05(50.0)	&5.02(50.0)\\  
 2	&1	&1	&1.5 &0.4	&4	&0.4	&4 &85.1(44.6)	&88.4(48.7)	&90.3(50.0)	&90.3(50.0)\\
3	&0.9	&1	&1.5  &0.38 	&3.9	&0.4	&4 &88.9(43.2)	&92.5(48.5)	&94.5(50.0)	&94.5(50.0)\\
4	&1.1	&1	&1.5  &0.42	&4.1	&0.4	&4 &78.7(43.3)	&80.6(48.1)	&81.4(50.0)	&81.5(50.0) \\
\hline
 \multicolumn{2}{c}{\textbf{Incongruent}} &&&&&&&&&&\\
5$^*$	&0	&1	&1	&0.2	&2.5	&0.4	&4 &8.6(0.7)	&7.3(1.1)	&7.4(0.0)	&7.3(0.0)\\
6$^*$	&-0.5 &1	&1	&0.14	&2	&0.4	&4 &6.3(0.0)	&6.7(0.0)	&7.3(0.0)	&7.3(0.0)\\
7	&2	&1	&1.5 &0.6	&5.5	&0.4	&4    &63.9(0.6)	&66.3(0.8)	&67.8(0.0)	&67.8(0.0)\\
8	&2.5	&1	&1.5  &0.7 	&6	&0.4	&4  &64.7(0.0) 	&66.3(0.0)	&67.8(0.0)	&67.8(0.0)\\
\hline
*Type I error
\end{tabular}}
\end{center}
\end{table}

\clearpage

\bigskip \bigskip \bigskip
\noindent\textit{K. Elastic power prior and elastic commensurate prior} \\
The idea of an elastic prior also can be applied to the power prior and commensurate prior, and we refer to them as elastic power prior and elastic commensurate prior.

\noindent\textit{K1. Elastic power prior} \\
With the power prior, the power parameter $\delta$ is treated as an unknown parameter, while with an elastic power prior, $\delta$ is linked with $T$ by an elastic function $g(\cdot)$, that is,
\begin{equation*}
\delta = g(T).
\end{equation*}
Then the elastic power prior is given by
\begin{equation*}
\pi^{*}(\theta | D_{h}) \propto \pi_{0}(\theta) f(D_{h} | \theta)^{g(T)},
\end{equation*}
where the elastic function $g(T)$ is same as logistic function (2) or step function (5) in Section 2, which maps support of $T$ to $[0, 1]$. The calibration of $g(T)$ is similar to that described in Section 2. Actually, the elastic power prior is identical to the calibrated power prior proposed by Pan \textit{et al} (2017).

Following the notations and initial priors described in Section 2, we display the elastic power prior for normal and binary endpoints. For the normal endpoint, the joint elastic power prior of  $(\theta, \sigma^2)$ is
\begin{equation*}
\begin{aligned}
\pi^{*}(\theta, \sigma^{2} | D_{h}) & \propto(\frac{1}{\sigma^{2}})^{m+\frac{g(T) n_{h}}{2}} \exp [-\frac{g(T) n_{h}}{2 \sigma^{2}}\{\widehat{\sigma}_{h}^{2}+(\theta-\overline{y}_{h})^{2}\}] \\ & \propto N_{\theta}( \overline{y}_{h}, \frac{\sigma^{2}}{g(T) n_{h}}) IG_{\sigma^2}( m+\frac{g(T) n_{h}-3}{2}, \frac{g(T) n_{h} \widehat{\sigma}_{h}^{2}}{2}).
\end{aligned}
\end{equation*}
Given current data, the posterior distribution for $(\theta, \sigma^2)$ is
\begin{equation*}
\pi( {\theta ,{\sigma ^2}|D,{D_h}} ) \propto N_{\theta}( {\frac{{g(T){n_h}{{\overline y }_h} + n_c\overline y_c }}{{g(T){n_h} + n_c}},\frac{{{\sigma ^2}}}{{g(T){n_h} + n_c}}}) IG_{\sigma^2}( {{\alpha ^\Delta },{\beta ^\Delta }}),
\end{equation*}
where ${\alpha ^\Delta } = m + \frac{{g(T){n_h} + n_c - 3}}{2}$,
${\beta ^\Delta } = \frac{{\sum\nolimits_{i = 1}^{n_c} {y_{c,i}^2 + g(T){n_h}\overline y _h^2} } + {g(T){n_h}\widehat \sigma _h^2} }{2} - \frac{{{{( {g(T){n_h}{{\overline y }_h} + n_c\overline y_c } )}^2}}}{{2{n_h}{g}(T) + 2n_c}}$.

For a binary endpoint, the elastic power prior of $p$ is
\begin{equation*}
\begin{aligned}
\pi^{*}(p | D_{h}) & \propto p^{g(T) n_{h} \overline{y}_{h}+\alpha_0-1}(1-p)^{g(T)(n_{h}-n_{h} \overline{y}_{h})+\beta_0-1} \\ & \propto Beta(g(T) n_{h} \overline{y}_{h}+\alpha_0, g(T)(n_{h}-n_{h} \overline{y}_{h})+\beta_0).
\end{aligned}
\end{equation*}
Based on the current data $D_c$, the posterior of $p$ is given as
\begin{equation*}
\pi(p | D, D_{h}) \propto Beta(g(T) n_{h} \overline{y}_{h}+\alpha_0 + n_c\overline{y}_c, g(T)(n_{h}-n_{h} \overline{y}_{h})+\beta_0 + n_c - n_c\overline{y}_c).
\end{equation*}

\noindent\textit{K2. Elastic commensurate prior} \\
With a commensurate prior, shrinkage parameter $\tau$ controls the degree that $\theta$ shrinks to $\theta_h$, and it is assumed unknown with a prior. However, with an elastic commensurate prior, $\tau$ is determined by $T$ through the elastic function $g(T)$, i.e.,
\begin{equation*}
\tau = g(T).
\end{equation*}
Then the elastic commensurate prior for $\theta$ is
\begin{equation*}
{\pi ^ * }( {\theta |{D_h},g(T)}) \propto \int_{{\theta _h}} {f( {{D_h}|{\theta _h}})\pi( {\theta |{\theta _h},g(T)}){\pi _0}( {{\theta _h}})} d{\theta _h}.
\end{equation*}
Since $\tau$ is located in $(0, +\infty)$, we adopt the following elastic function:
\begin{equation*}
g(T) = exp(a + b\cdot \log(T)).
\end{equation*}
If a larger value of $T$ indicates more incongruence between $D_c$ and $D_h$, we require $b < 0$ to ensure that a larger value of $T$ leads to a smaller value of $g(T)$ (i.e., a larger variance inflation). The calibration of $g(T)$ is similar to that described in Section 2.

Let us return to the Gaussian case. We first focus on the historical information borrowing for location parameter $\theta$, that is $\theta|\theta_h \sim N(\theta_h,\tau^{-1} )$, where $\tau=g(T)$. Assuming $\pi_0(\theta_h) \propto 1$ and integrating out the nuisance parameter $\theta_h$, the elastic commensurate prior for $\theta$ is
\begin{equation*}
\pi^\ast(\theta|D_h, g(T))\propto N({\overline{y}}_h,\frac{1}{g(T)}+\frac{{\widehat{\sigma}}_h^2}{n_h}).
\end{equation*}
Multiplying the above elastic commensurate prior with the current likelihood, we obtain the following posterior distribution for $\theta$:
\begin{equation*}
\pi( {\theta |D,{D_h},{\sigma ^2}}) \propto N( {\frac{{n_c\overline y_c \Delta  + {\sigma ^2}{{\overline y }_h}}}{{n_c{\Delta} + {\sigma ^2}}},\frac{{{\sigma ^2}{\Delta}}}{{n_c{\Delta} + {\sigma ^2}}}}),
\end{equation*}
where $\Delta  = \frac{1}{{g(T)}} + \frac{{\widehat \sigma _h^2}}{{{n_h}}}$.

If the information borrowing both for location parameter $\theta$ and scale parameter $\sigma^2$ are required, a new precision parameter $\zeta$ is introduced to measure the commensurate between $\sigma^2$ and $\sigma_h^2$. Specifically, we assume $\sigma^2$ a prior that is centered at $\sigma_h^2$ with precision $\zeta$, e.g.,
$ \sigma ^2| \sigma_h^2 \sim IG( {\sigma _h^2,{\zeta ^{ - 1}}})$, where $IG(\cdot)$ is an inverse gamma distribution with mean $\sigma _h^2$ and variance $\zeta ^{ - 1}$. With an elastic commensurate prior, precision $\tau=g_1(T_1)$ and $\zeta=g_2(T_2)$. Given historical data $D_h$, assuming a prior $\pi_0(\sigma_h^2) \propto (\sigma_h^2)^{-m}$ for $\sigma_h^2$ and integrating out $\theta_h$, the joint elastic commensurate prior for $(\theta, \sigma^2)$ is
\begin{equation*}
\begin{aligned}
{\pi ^ * }( {\theta ,{\sigma ^2},\sigma _h^2|{D_h},{g_1}(T_1),{g_2}(T_2)}) &\propto f(D_h|\theta_h, \sigma_h^2)N_{\theta}(\theta_h, g_1(T_1)^{-1})IG_{\sigma^2}(\alpha^{'}, \beta^{'})\times (\sigma_h^2)^{-m} \\
&\propto N_{\theta }( {{\overline y }_h},\frac{1}{{g_1}(T_1)} + \frac{\sigma _h^2}{n_h})IG_{\sigma ^2}( \alpha^{'} ,\beta^{'} )\\
 &\times IG_{\sigma _h^2}( {\frac{{{n_h} - 3}}{2} + m,\frac{{{n_h}\widehat \sigma _h^2}}{2}}),
\end{aligned}
\end{equation*}
where $\alpha^{'}=g_2(T_2)\sigma_h^4 + 2$ and $\beta^{'}=\sigma_h^2(g_2(T_2)\sigma_h^4+1)$.

\end{document}